\documentclass[10pt]{article}

\usepackage{amsmath}
\usepackage{amssymb}

\usepackage[dvipdfmx]{graphicx}
\usepackage{comment}

\usepackage{cite}

\usepackage{color} 

\usepackage{setspace}

\usepackage[labelfont=bf,labelsep=period,justification=raggedright]{caption}

\makeatletter
\renewcommand{\@biblabel}[1]{\quad#1.}
\makeatother

\date{}

\begin{document}

% Title must be 150 characters or less
\begin{flushleft}
{\Large
\textbf{Invasion of cooperation in scale-free networks: Accumulated vs.\ average payoffs}
}
% Insert Author names, affiliations and corresponding author email.
\\
Genki Ichinose$^{1\ast}$, 
Hiroki Sayama$^{2}$, 
\\
\bf{1} Department of Mathematical and Systems Engineering, Shizuoka University, Hamamatsu, 432-8561, Japan\\
\bf{2} Center for Collective Dynamics of Complex Systems, Binghamton University, State University of New York, Binghamton, NY 13902-6000, USA
\\
$\ast$ E-mail: ichinose.genki@shizuoka.ac.jp
\end{flushleft}

% Please keep the abstract between 250 and 300 words
\section*{Abstract}
It is well known that cooperation cannot be an evolutionary stable strategy for a non-iterative game in a well-mixed
population.
In contrast, structured populations favor cooperation since cooperators can benefit each other by forming
local clusters.
Previous studies have shown that scale-free networks strongly promote cooperation.
However, little is known about the invasion mechanism of cooperation in scale-free networks.
To study microscopic and macroscopic behaviors of cooperators' invasion, we conducted computational experiments of the evolution of cooperation in scale-free networks
where, starting from all defectors, cooperators can spontaneously emerge by mutation.
Since the evolutionary dynamics are influenced by the definition of fitness, we tested two commonly adopted fitness functions: accumulated payoff and average payoff.
Simulation results show that cooperation is strongly enhanced with the accumulated payoff fitness compared to the average payoff fitness.
However, the difference between the two functions decreases as the average degree increases.
As the average degree increases, cooperation decreases with the accumulated payoff fitness, while it increases with the average payoff fitness.
Moreover, with the average payoff fitness, low-degree nodes play a more important role in spreading cooperative strategies compared to the case of the accumulated payoff fitness.

\section*{Keywords}
Evolution of cooperation, Prisoner's Dilemma, evolutionary game, scale-free network, invasion dynamics, fitness function

\section*{Authors' information}
\begin{itemize}
\item Dr. Genki Ichinose (Corresponding author)\\
Affiliation: Department of Mathematical and Systems Engineering, Shizuoka University\\
Address: 3-5-1 Johoku, Naka-ku, Hamamatsu City, 432-8561, Japan\\
E-mail: ichinose.genki@shizuoka.ac.jp\\
Tel: +81-53-478-1211
\item Dr. Hiroki Sayama\\
Affiliation: Center for Collective Dynamics of Complex Systems, Binghamton University, State University of New York\\
Address: P.O. Box 6000, Binghamton, NY 13902-6000, USA\\
E-mail: sayama@binghamton.edu\\
Tel: +1-607-777-3566
\end{itemize}
% Please keep the Author Summary between 150 and 200 words
% Use first person. PLoS ONE authors please skip this step. 
% Author Summary not valid for PLoS ONE submissions.   

\newpage
\doublespacing
\section*{Introduction}
The emergence of cooperation is one of the challenging problems in
both the biological and social sciences.  Cooperators benefit others by
incurring some costs to themselves while defectors do not pay any
costs. Therefore, cooperation cannot be an evolutionary stable
strategy for a non-iterative game in a well-mixed population.
This relationship between cooperators and defectors is well parameterized
in the Prisoner's Dilemma game (PD) \cite{Axelrod1984}.
In PD, two individuals decide whether to cooperate or defect simultaneously.
They both obtain $R$ for mutual cooperation or $P$ for mutual defection as payoffs.
If one selects cooperation and the other selects defection,
the former receives $S$ for being the ``sucker'' of the defection,
while the latter receives $T$ as a reward, which is the ``temptation'' to defect.
The order of the four payoffs is $T > R > P > S$ in PD.

Nowak and May were the first to reveal that spatial structure provides a viable mechanism for cooperation to evolve \cite{NowakMay1992}.
Recently, spatial structures have been mapped to suitable network topologies
and the evolution of cooperation has been investigated
through the analysis of PD played
on those network topologies
\cite{AbramsonKuperman2001, MasudaAihara2003, Masuda2007,
Ohtsuki_etal2006, Ashlock2007, SantosPacheco2005, IchinoseKobayashi2011, Ichinose_etal2013}.
 In this context, the effect of spatial structures required for the emergence of cooperation is referred
to as network reciprocity, which has come to be recognized as one of the enabling mechanisms
for the emergence of cooperation as well as the other ones \cite{Nowak2006}.

On such spatial structures, cooperators can form clusters and thereby reduce the risk of exploitation by defectors.
In particular, it has been reported that scale-free networks strongly promote the evolution of cooperation \cite{SantosPacheco2005}.
If a cluster of cooperators takes over a hub in a scale-free network,
the payoffs for these cooperators are considerably higher than those for other individuals, and thus
they can spread their cooperative strategy quickly to the entire network.
In contrast, if a cluster of defectors takes over a hub,
this cluster of defectors is quite vulnerable and is easily replaced by cooperators.
These effects explain why cooperation is likely to evolve on scale-free networks.
 
However, it is still unknown how cooperator-dominated hubs could arise in society that initially has few cooperators.
It was assumed in \cite{SantosPacheco2005} that the initial state of the social network was made of half cooperators and half defectors.
 This setting has since been adopted by other studies on evolutionary models of cooperation as described in \cite{PercSzolnoki2010, PercGrigolini2013}.
This assumption does not explain how one cooperator that emerged by mutation could increase in number and invade into the population initially filled by defectors.
This invasion dynamics has already been investigated for square lattice network topologies \cite{Langer_etal2008, Fu_etal2010}.
In contrast, little is known about such invasion dynamics on scale-free networks, although the fixation probability on scale-free networks has been reported in \cite{Ohtsuki_etal2006, Li_etal2013}.
Recently, Miller and Knowles studied a model with a similar premise where coevolution of strategy and network growth starting from a very small population is considered \cite{MillerKnowles2016, MillerKnowles2016b}.
Pinheiro et al. developed a ``gradient of selection'', which numerically decides whether cooperation increases or decreases under given fraction of cooperators (including only one cooperator in a population) \cite{Pinheiro_etal2012a, Pinheiro_etal2012b}.
However, the microscopic mechanism of cooperators' invasion is not obvious in those studies because the analysis only focuses on fitness difference among all individuals without taking into account how and where cooperation invades into a network.

Moreover, there is room for further investigation about how the evolution of cooperation depends on the accumulated payoff fitness, which is usually assumed in the previous studies mentioned above.
The primary reason that cooperation is strongly promoted in scale-free networks is that cooperative hubs can gain extremely high payoffs compared to the other nodes because such hubs have a large number of connections to other cooperators.
This effect of degree heterogeneity disappears if payoffs are averaged. Therefore, in the case that averaged payoffs are used \cite{SantosPacheco2006, Tomassini_etal2007, Wu_etal2007, Szolnoki_etal2008} or some costs are incurred to maintain links \cite{Masuda2007}, the evolution of cooperation is strongly inhibited.
In this way, the evolutionary dynamics are greatly influenced by the definition of fitness.
The key question to be investigated is whether cooperation emerges from such harsh environment and how it spreads into a network by analyzing the microscopic behaviors of strategy evolution.

 In this paper, we performed computer simulations of the evolution of cooperation in scale-free networks
 where the initial population is all defectors but cooperators can spontaneously arise by mutation.
 In these simulations, we tested two commonly adopted fitness functions: accumulated payoff and average payoff, while the average degree of nodes is systematically varied.
 The purpose of this study is to reveal the microscopic and macroscopic behaviors of the cooperators' invasion into the network, and how they differ between the two fitness assumptions.

% You may title this section "Methods" or "Models". 
% "Models" is not a valid title for PLoS ONE authors. However, PLoS ONE
% authors may use "Analysis" 
\section*{Models}
 We consider the evolutionary dynamics of cooperators' invasion in scale-free networks.
 The Barab\'{a}si-Albert method is used for generating initial networks in simulations \cite{BarabasiAlbert1999}.
 Then, each generated network is substantially randomized by the double-edge swap method while
keeping the original degree distributions in order to remove artificial network properties that are known to occur in the BA model \cite{Bollobas_etal2003, FotouhiRabbat2013}.
Such randomized scale-free networks are also used in \cite{SantosPacheco2006} and it is known that, in this case, cooperation is inhibited a little compared to the original BA scale-free network.
 Self loops and multiedges are avoided during the randomization.

 A network is made of $N$ nodes occupied by individuals.
 Each node has its strategy classified as either C (cooperator) or D (defector).
 Initially, all individuals are defectors. Each node $i$ plays the Prisoner's Dilemma game (PD) with all of its $k_i$ neighbors.
 The payoffs of the game are calculated as follows.
 Both individuals obtain $R$ for mutual cooperation and $P$ for mutual defection.
 If one selects cooperation and the other selects defection, the cooperator obtains $S$ as the sucker of defection, and the defector obtains $T$ as the reward for temptation to defect.
 The order of the four payoffs is $T > R > P \geq S$ in typical PD. In the case that $P=S$, the game is called weak prisoner's dilemma.
Following previous studies \cite{NowakMay1992, SantosPacheco2005}, we set $P = 0$, $T = b$, $R = 1$, and $S = 0$, where $b > 1$ is the only control parameter.
 The payoff of individual $i$ against its $k_i$ neighbors is denoted by $p_i$.
 Here we consider two types of $p_i$: accumulated payoff and average payoff.
 The average payoff is obtained by dividing the accumulated payoff by $k_i$.
 
We assume an asynchronous updating in our model as used in other recent models of evolutionary games where the following operations are applied to each individual.
 At the beginning of each operation, one randomly selected individual $x$ plays PD with its neighbors and obtains payoff $p_x$.
 Next, one randomly chosen neighbor of $x$, denoted by $y$, also plays PD with its neighbors and obtains payoff $p_y$.
 If $p_x < p_y$, individual $x$ imitates individual $y$'s strategy with probability  $(p_y - p_x)/[(T - S)k_{\mathrm{max}}]$ \cite{SantosPacheco2005}, where $k_{\mathrm{max}} = \mathrm{max}(k_x,k_y)${, for the accumulated payoff condition and $(p_y - p_x)/(T - S)$, for the average payoff condition.
 $k_{\mathrm{max}}$ is used for normalization to be the probability less or equal than 1.
 Finally, another randomly selected individual $z$ ($z$ might be the same as $x$ or $y$) flips its strategy (C will become D and D will become C) by mutation with probability $m$.
 These operations consist of the one time step (usually called ``Monte Carlo step'' \cite{PercSzolnoki2010}).
 
 We regard $N$ time steps as one generation, in which all individuals are selected once, on average, for the strategy update
and mutation.

% Results and Discussion can be combined.
\section*{Results}
\subsection*{Macroscopic dynamics of cooperator invasion}
%\subsubsection*{Fraction of cooperators}
First, we focus on macroscopic dynamics of the cooperators' invasion.
We compared simulation results for the two fitness conditions (accumulated and average payoff fitnesses).
For each fitness condition, we conducted simulations by varying two parameters: the temptation to defect ($b$, from 1.1 to 2.0) and the average node degree ($\bar{k}$, from 4 to 16). The same tendency of the results holds for the larger two values.
The other parameters used in the simulations were $N=5,000$ (population size) and $m=0.005$ (mutation probability).
Results are shown in Fig.~\ref{fracCdegree}, which clearly shows the effects of $b$ and $\bar{k}$, as well as the difference between the two fitness conditions.

Basically, cooperation is greatly enhanced in the accumulated payoff fitness condition than in the average payoff fitness condition, as previous studies have reported \cite{SantosPacheco2005, SantosPacheco2006}\footnote{However, the final levels of cooperation are lower in our model than in those other models, because there are no cooperators in initial configurations and the clusters of cooperators can collapse due to mutation in our model.}.
This is because cooperative hubs can gain much greater payoffs than others by having a large number of connections to other cooperators, as long as the temptation to defect $b$ is not too large.
However, our results show that, if $b$ is large (e.g., $b>1.8$, $\bar{k}=8$), cooperators may not be able to occupy hub nodes and
therefore the cooperation is not promoted any more.

Moreover, in the accumulated payoff fitness condition, cooperation {\em decreases} as the average degree increases, which disagrees with what was originally reported in \cite{SantosPacheco2005}.
This disagreement is due to the difference in model settings: In the model used in \cite{SantosPacheco2005}, the initial population was filled by half cooperators and half defectors, so having a higher average degree made it easier for cooperators to form a cluster initially by themselves.
Once such clusters are formed, they hardly collapse because there is no mutation in \cite{SantosPacheco2005}.
In contrast, our model assumes that the initial condition is full of defectors, and that cooperators appear only by mutation. In this model setting, it is easier for cooperative clusters to form if the average degree is low, because a cooperator will be connected to fewer defectors.
We also checked the usual case in which the initial population is filled by half cooperators and half defectors and found that the difference among the average degrees are weakened in such a case (results not shown).

In contrast, cooperation {\em increases} as the average degree also increases in the average payoff fitness condition. Therefore, the difference between the two functions decreases as the average degree increases.
This is because a probability of a rare cluster of cooperators being able to connect to other cooperators increases as the average degree increases. In that case, the chance for cooperators to survive increases.
Note that this situation happens only when cooperation is not dominant such as the initial stage of invasion. Once cooperation prevails, defectors can exploit cooperators more as the average degree increases because a probability of being connected to cooperators increases in such a situation.
Then, the situation approaches a well-mixed population, which is harmful for cooperation, inversely.

In brief, our results showed an interesting difference between the two fitness conditions, in terms of the effect of average node degrees. 
These findings rely on our unique model settings in which cooperators are initially non-existent and they spontaneously arise by mutation.

\subsection*{Microscopic dynamics of cooperators' invasion}
%\subsubsection*{Difference of strategy expansion between cooperators and defectors}
Next, we investigate the microscopic dynamics of cooperator invasion.
We used $\bar{k}=8$ and $b=1.2$ as a representative parameter setting throughout this section, because the general trend was consistent even if they are varied.
Figure \ref{SumSrcDstDegree} shows histograms of strategy propagation events, plotted over
the degrees of source and destination nodes.
Spontaneously emerged cooperators by mutation try to invade a network of defectors.
In that phase, if clusters of cooperators are formed by chance, the invasion is likely to succeed.
This situation happens in the first 300-400 generations as you can see in Fig.~\ref{FracCvsTime}.
After that, only minor changes take place.
Therefore, in each case, the first 500 generations of 10 replicate simulations are recorded because the first 500 generations are enough to see the invasion of cooperation. 

As seen in Fig.~\ref{SumSrcDstDegree}A and B, only lower-degree nodes can change higher-degree nodes' strategies in the average payoff fitness condition.
As a result, cooperation tends to spread more frequently from lower degree nodes in the average payoff fitness condition (Fig.~\ref{SumSrcDstDegree}A) than in the accumulated payoff fitness condition (Fig.~\ref{SumSrcDstDegree}C).
This is because the benefit of being hubs for cooperators disappears in the average payoff fitness condition, as discussed above.
In contrast, a relatively wide range of node degrees can cause a change in the strategy of neighbors although the strategy of hub nodes cannot be changed in the accumulated payoff fitness condition (Fig.~\ref{SumSrcDstDegree}C and D).

To investigate the effects of the local surrounding environment for the propagation of cooperation, we also plotted histograms of strategy propagation events over
the degree of the source node and its neighbors' state ratio (1 = fully cooperative neighborhood, 0 = fully defective neighborhood) in Fig.~\ref{CumDegreeState}.
In the accumulated payoff fitness condition, a node with any degree can change its neighbors' strategy in general. Moreover, as neighbors' state ratio becomes greater, the frequency of strategy change tends to be greater because such a high cooperation ratio contributes to raise the fitness of the source's node.
In contrast, in the average payoff fitness condition, only low node degrees with any neighbors' ratio tend to be a cause of change in the strategy of neighbors as shown in Fig.~\ref{CumDegreeState}A.
In the case of high source node degrees in Fig.~\ref{CumDegreeState}A, cooperators are easily invaded by defectors because defectors simply can get high payoff from such a high ratio of cooperators, resulting in no existence of defectors with a high neighbor ratio of cooperators as shown in Fig.~\ref{CumDegreeState}B.

\subsection*{Case of donation game}
In all the simulations above, we adopted a ``weak'' PD setting where the sucker's payoff ($S$) is equal to the punishment ($P$), which is a common assumption made in many earlier studies (e.g., \cite{NowakMay1992, SantosPacheco2005}).
However, this assumption does not create an incentive for cooperators to switch their strategy to defection when they play a game with defectors.

In order to test the robustness of our findings in a ``strong'' PD setting with $T>R>P>S$ and $2R > T+S$, we conducted another set of simulations using the ``donation game'' model. The donation game is a special class of ``strong'' PD where
each cooperator provides a benefit $b$ to the other player by incurring cost $c$ to himself, with $0<c<b$.
Thus, the payoff structure of the donation game is given by $T=b$, $R=b-c$, $P=0$, and $S=-c$ \cite{Hilbe_etal2013}.
For simplicity, we varied just one parameter $b$ from 1 to 2, while letting $c = b - 1$.
Figure \ref{fracCdonation} shows the fraction of cooperators on scale-free networks in the simulations of the donation game.
We find cooperation is greatly inhibited in the donation game compared to the weak PD.
Cooperation sharply drops in both cases and the same tendency between average and accumulated payoff fitness conditions with the weak PD can only be seen in the low limited range of $b$ ($b=1.1$).
In the case of the weak PD, the payoffs between $P$ and $S$ are the same by definition, making cooperators not so disadvantageous against defectors.
In contrast, in the donation game, defectors can easily exploit cooperators.
Thus, cooperation can only survive within the limited range of $b$.

\section*{Conclusion}
It is known that scale-free networks strongly promote cooperation due to their heterogeneity.
It is also known that this advantage is mostly lost when average payoffs are adopted.
However, little is known about how cooperation can spread from initially rare cooperators in the two cases.
Here we show that the fate of cooperation between the average and accumulated payoff fitness condition is very different depending on the average degree.
In general, cooperation is promoted in the accumulated payoff fitness condition as previously found.
However, the difference between accumulated and average payoff decreases as the average degree increases, which is not commonly discussed in the previous studies.
More importantly, as the average degree increases, cooperation decreases with the accumulated payoff fitness, while it increases with the average payoff fitness.
This implies that the evolution of cooperation on a network depends significantly on how game players are rewarded through their game play.
Moreover, from the in-depth analysis of microscopic behaviors, we show that the relative importance of low degree nodes for the evolution of cooperation is much higher in the case of the average payoff,
compared to the previously studied case of the accumulated payoff where hubs are known to play a major role in the propagation of cooperation.

% Do NOT remove this, even if you are not including acknowledgments
\section*{Acknowledgments}
The authors thank Yoshiki Satotani for his comments on this work.

%\section*{References}
% The bibtex filename
%\bibliography{PLoS_com}
\providecommand{\noopsort}[1]{}\providecommand{\singleletter}[1]{#1}%

\newpage
\section*{Figures}
%\begin{figure}[!ht]
%\begin{center}
%%\includegraphics[width=4in]{figure_name.2.eps}
%\end{center}
%\caption{
%{\bf Bold the first sentence.}  Rest of figure 2  caption.  Caption 
%should be left justified, as specified by the options to the caption 
%package.
%}
%\label{Figure_label}
%\end{figure}

\begin{figure}[htbp]
\begin{center}
\includegraphics[width=\textwidth]{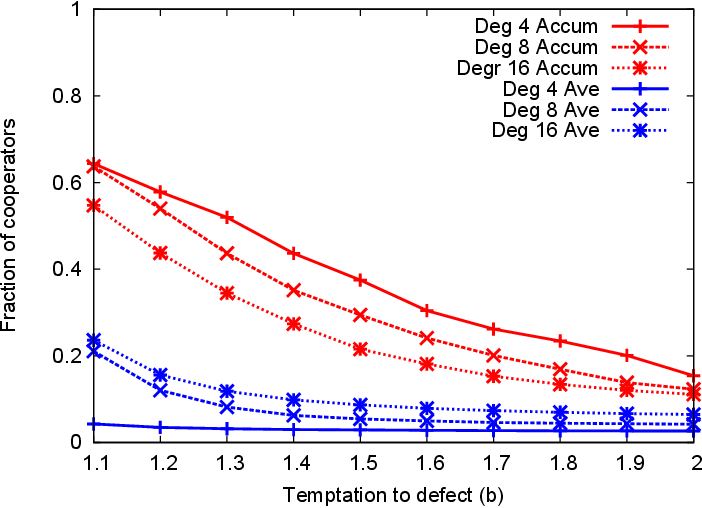}
\caption{Fraction of cooperators against $b$ in different average degree settings. The average of 10 independent simulation
runs (the last 1,000 generations for each) is shown.}
\label{fracCdegree}
\end{center}
\end{figure}

\begin{figure}[htbp]
\begin{center}
\includegraphics[width=\textwidth]{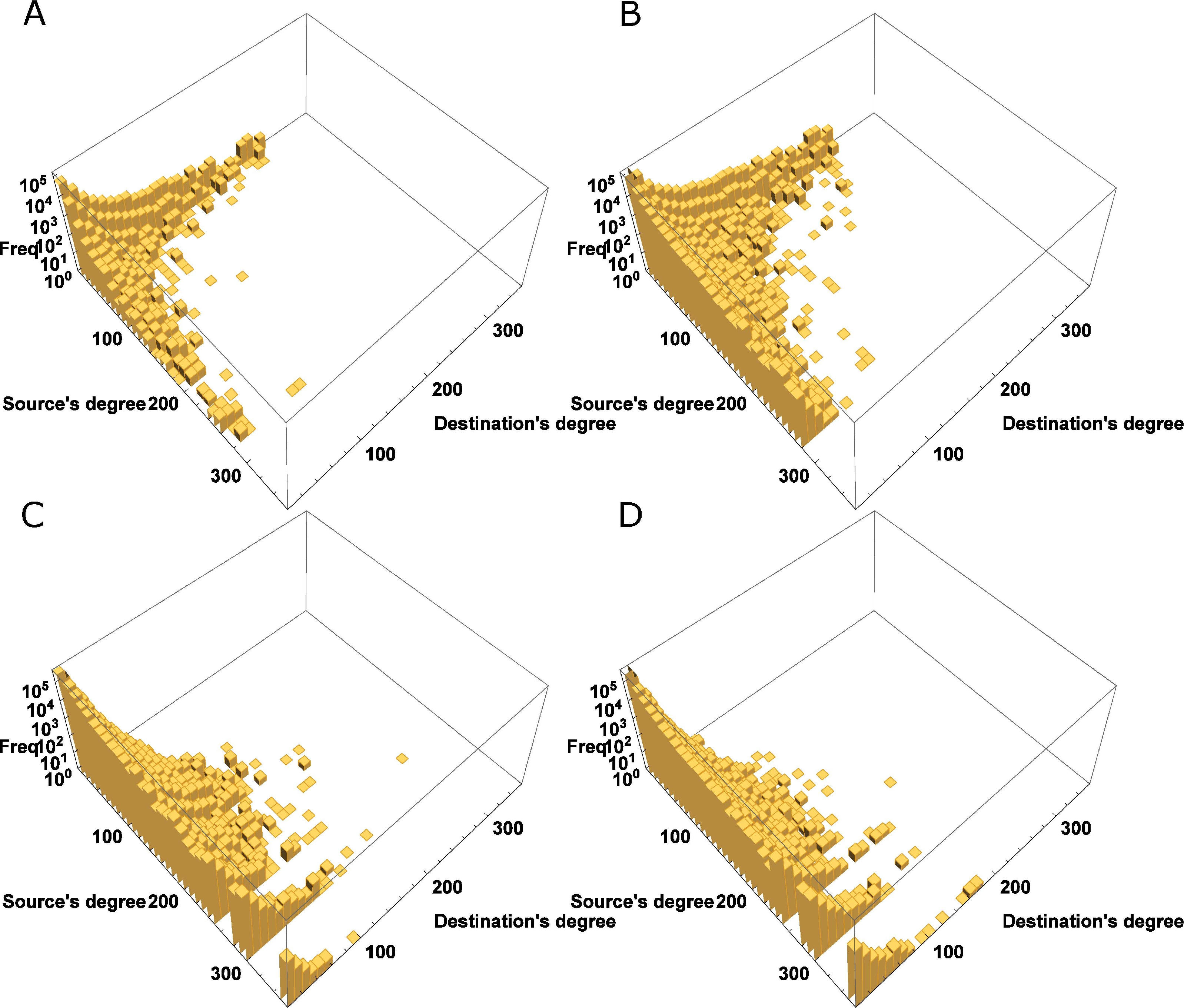}
\caption{Frequency of strategy propagation events, plotted over the source node's degree
and the destination node's degree in the average payoff fitness condition (top: A and B) and the accumulated payoff fitness condition (bottom: C and D). Cooperation propagation is on the left (A and C) and defection propagation is on the right (B and D).
$N=5,000$, $\bar{k}=8$, $m=0.005$ , and $b=1.2$. First 500 generations of 10 replicate simulations are recorded.}
\label{SumSrcDstDegree}
\end{center}
\end{figure}

\begin{figure}[htbp]
\begin{center}
\includegraphics[width=\textwidth]{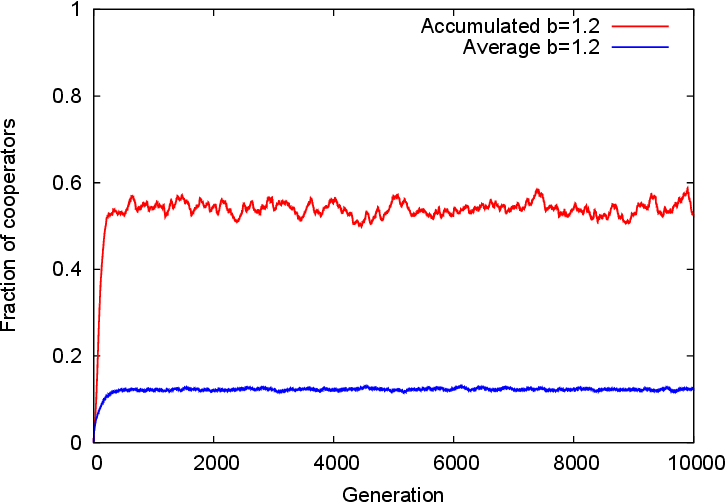}
\caption{Fraction of cooperators vs. time. The average of 20 independent simulation runs is shown. $N=5,000$, $\bar{k}=8$, $m=0.005$, and $b=1.2$.}
\label{FracCvsTime}
\end{center}
\end{figure}

\begin{figure}[htbp]
\begin{center}
\includegraphics[width=\textwidth]{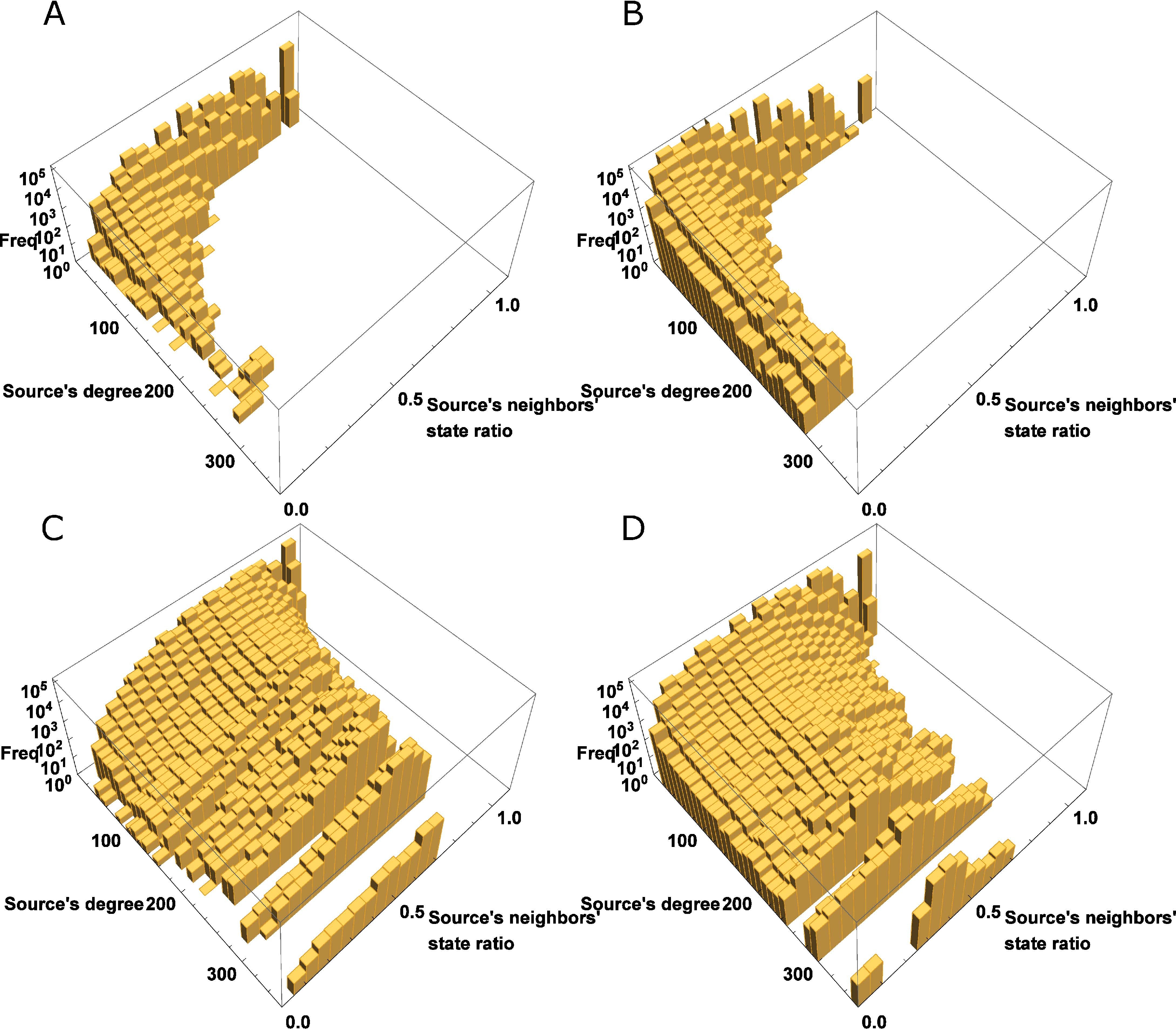}
\caption{Frequency of strategy propagation events, plotted over the source node's degree
and its neighbors' state ratio in the average payoff fitness condition (top: A and B) and the accumulated payoff fitness condition (bottom: C and D).
Cooperation propagation is on the left (A and C) and defection propagation is on the right (B and D).
$N=5,000$, $\bar{k}=8$, $m=0.005$, and $b=1.2$.
First 500 generations of 10 replicate simulations are recorded.}
\label{CumDegreeState}
\end{center}
\end{figure}

\begin{figure}[htbp]
\begin{center}
\includegraphics[width=\textwidth, clip]{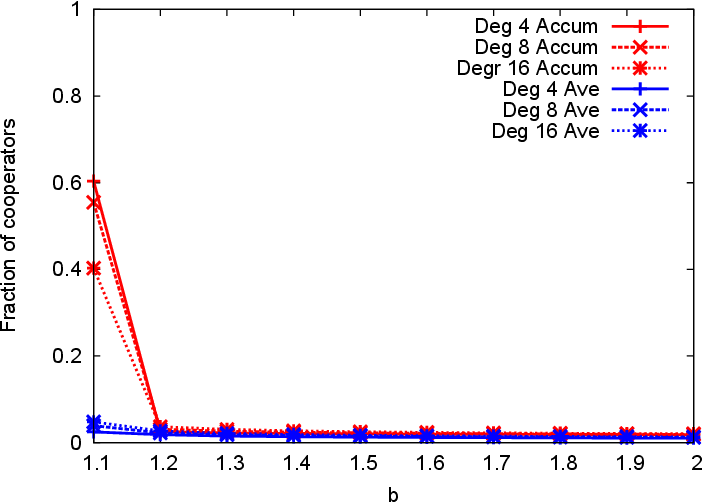}
\caption{Fraction of cooperators in the case of the donation game.} The average of 10 independent simulation
runs (the last 1,000 generations for each) is shown.
\label{fracCdonation}
\end{center}
\end{figure}

%\section*{Tables}
%\begin{table}[!ht]
%\caption{
%\bf{Table title}}
%\begin{tabular}{|c|c|c|}
%table information
%\end{tabular}
%\begin{flushleft}Table caption
%\end{flushleft}
%\label{tab:label}
% \end{table}

\end{document}